# Reciprocal Symmetry and the Origin of Spin

Mushfiq Ahmad
Department of Physics, Rajshahi University, Rajshahi, Bangladesh.
E-mail: mushfiqahmad@ru.ac.bd

M. Shah Alam
Department of Physics, Shah Jalal University of Science and Technology, Sylhet, Bangladesh. E-mail: salam@sust.edu

## Abstract

We have shown that Reciprocal Symmetric transformation shares the algebraic properties of Dirac Electron Theory more than Lorentz transformation and that the origin of spin is in Reciprocal Symmetry.

## 1. Introduction

Spin appears in Dirac Electron Theory, which is based on Lorentz invariance of total energy. This will inspires one to look for spin in Lorentz transformation. Although Lorentz transformation has some rotational properties[1], but no spin or any involvement with Pauli matrices is found in it. Reciprocal Symmetric transformation[2] also fulfills Lorentz invariance requirement. Naturally, Reciprocal Symmetry[3] is the next place to look for spin.

## 2. 4-Diemnsional Vector

We construct a 4-dimensional vector $\mathsf{A}$. We follow the convention adopted by Kyrala[4] to write it as a sum of a scalar, $a_0$, and a 3-dimmensioanl Cartesian vector, $\mathbf{A} = \mathbf{i}a_1 + \mathbf{j}a_2 + \mathbf{k}a_3$. We include the scalar as the $0^{th}$ component.

$$\mathsf{A} = a_0 + \mathbf{A} \text{ and } \mathsf{B} = (b_0 + \mathbf{B}) \tag{1}$$

We also define the conjugate

$$\mathsf{A}^* = a_0 - \mathbf{A} \tag{2}$$

and the norm

$$|\mathsf{A}| = \left|\sqrt{a_0^2 + \mathbf{A}.\mathbf{A}}\right| \tag{3}$$

## 3. Lorentz-Einstein Rule of Composition

We define a Lorentz-Einstein rule of composition, $<\times>$, as

$$\mathsf{A} <\times> \mathsf{B} = a_0 b_0 + \mathbf{A}.\mathbf{B} + \mathbf{A}\sqrt{b_0^2 - \mathbf{B}^2} + \left\{\left[b_0 - \sqrt{b_0 - \mathbf{B}^2}\right]\frac{\mathbf{A}.\mathbf{B}}{\mathbf{B}^2} + a_0\right\}\mathbf{B} \tag{4}$$

Lorentz-Einstein rule of composition gives

$$\mathsf{A}^2 = \mathsf{A} <\times> \mathsf{A}^* = a_0{}^2 - \mathbf{A}.\mathbf{A} \tag{5}$$

(4) also gives

$$(\mathsf{A} <\times> \mathsf{B})^2 = \mathsf{A}^2\mathsf{B}^2 \tag{6}$$

## 4. Reciprocal Symmetric Rule of Composition

We define a reciprocal symmetric rule of composition, $(\times)$, as

$$\mathsf{A}(\times)\mathsf{B} = a_0 b_0 + \mathbf{A}.\mathbf{B} + b_0\mathbf{A} + a_0\mathbf{B} + i\mathbf{A}\mathrm{x}\mathbf{B} \tag{7}$$

Reciprocal symmetric rule of composition gives

$$\mathsf{A}^2 = \mathsf{A}(\times)\mathsf{A}^* = a_0{}^2 - \mathbf{A}.\mathbf{A} \tag{8}$$

(7) also gives

$$(A(x)B)^2 = A^2B^2 \tag{9}$$

## 5. Lorentz Invariance

If we set

$$b_0 = \frac{c}{\sqrt{c^2 - \mathbf{V}^2}} \quad \text{and} \quad \mathbf{B} = \frac{\mathbf{V}}{\sqrt{c^2 - \mathbf{V}^2}} \tag{10}$$

(5) and (7) give

$$B^2 = 1 \tag{11}$$

In this case (6) and (9) give

$$(A < x > B)^2 = A^2 \text{ and } (A(x)B)^2 = A^2 \tag{12}$$

(12) are Lorentz invariance relations.

## 6. Pauli Quaternion

To study the relation of Reciprocal symmetric rule of composition to Pauli matrix[5] algebra, from 4-vector $A$ we construct the Pauli Quaternion[6], $\hat{A}$

$$\hat{A} = \sigma_0 a_0 + \sigma_x a_x + \sigma_y a_y + \sigma_z a_z = \sigma_0 A_0 + \boldsymbol{\sigma}.\mathbf{A} \tag{13}$$

Basis vectors $\sigma_0$, $\sigma_x$ etc. have the properties of Pauli matrices[7]

$$\sigma_0^2 = \sigma_x^2 = \sigma_y^2 = \sigma_z^2 = 1 \tag{14}$$

$$\sigma_x\sigma_0 - \sigma_0\sigma_x = \sigma_y\sigma_0 - \sigma_0\sigma_y = \sigma_z\sigma_0 - \sigma_0\sigma_z = 0 \tag{15}$$

$$\sigma_x\sigma_y + \sigma_y\sigma_x = \sigma_x\sigma_z + \sigma_z\sigma_x = \sigma_y\sigma_z + \sigma_z\sigma_y = 0 \tag{16}$$

And

$$\sigma_x\sigma_y = i\sigma_z,\ \sigma_y\sigma_z = i\sigma_x,\ \sigma_z\sigma_x = i\sigma_y \tag{17}$$

The direct product gives

$$\hat{A}\hat{B} = \sigma_0(a_0 b_0 + \mathbf{A}.\mathbf{B}) + \boldsymbol{\sigma}.(b_0\mathbf{A} + a_0\mathbf{B} + i\mathbf{A}\mathrm{x}\mathbf{B}) \tag{18}$$

Therefore, Reciprocal symmetric rule of composition (7) agrees with Pauli matrix algebra.

## 7. Dirac Algebra

We now take a look at Dirac algebra[8]. We start with the wave equation

$$(E - c\boldsymbol{\alpha}.\mathbf{p} - \beta mc^2)\psi = 0 \tag{19}$$

For the purpose of comparison we set $m = 0$. This makes it possible to replace $\boldsymbol{\alpha}$ by $\boldsymbol{\sigma}$ and we have

$$(\sigma_0 E - \boldsymbol{\sigma}.p)\psi = 0 \tag{20}$$

$\sigma_0 E - \boldsymbol{\sigma}.p$ corresponds to

$$\hat{A} = \sigma_0 a_0 + \boldsymbol{\sigma}.\mathbf{A} \tag{21}$$

## 8. Reciporcal Symmetry and Spin

In the presence of electro magnetic field one of the term Dirac theory gives is (using the notation of Schiff[9]).

$$(\boldsymbol{\alpha}.\mathbf{B})(\boldsymbol{\alpha}.\mathbf{C}) = \mathbf{B}.\mathbf{C} + i\boldsymbol{\alpha}.(\mathbf{B}\mathrm{x}\mathbf{C}) \tag{22}$$

The cross term gives spin. Comparison with (7) and (18) shows that Reciprocal Symmetric algebra laso give this cross term. Therefore, we see that the origin of spin is in Reflection Symmetry.

## 9. Conclusion

Spin is embedded in the algebra of reciprocal symmetric transformation. Lorentz transformation has different algebraic properties and does not contain spin.

---

[1]Mushfiq Ahmad, M. Shah Alam, M.O.G. Talukder. Comparison between Spin and Rotation Properties of Lorentz Einstein and Reflection Symmetric Transformations. http://www.arxiv.org/find

[2] Md. Shah Alam and M.H. Ahsan - Mixed Number Lorentz Transformation.Vol16. No4. 2003. Physics Essays.

[3] Mushfiq Ahmad. Reciprocal Symmetry and Equivalence between Relativistic and Quantum Mechanical Concepts. http://www.arxiv.org/find

[4]A. Kyrala. Theoretical Physics: Applications of Vectors, Matrices, Tensors and Quaternions. W. B. Saunders Company, Philadelphia & London. 1967.

[5]Schiff, L.I (1970). Quantum Mechanics. Mc Graw-Hill Company. Third Edition.

[6] George Raetz. http://home.pcisys.net/~bestwork.1/quaterni2.html

[7] Schiff, L.I (1970). Quantum Mechanics. Mc Graw-Hill Company. Third Edition.

[8] Schiff, L.I (1970). Quantum Mechanics. Mc Graw-Hill Company. Third Edition.

[9] Schiff, L.I (1970). Quantum Mechanics. Mc Graw-Hill Company. Third Edition.